\documentstyle[12pt,psfig,epsf]{article}
\def\href#1#2{#2}
\begin{document}

\thispagestyle{empty}
 \begin{flushleft}
DESY 97--219
\\
hep-ph/9711439
\\
November 1997
 \end{flushleft}

% DESY 97-219
%
 \noindent
 \vspace*{0.50cm}
\begin{center}
 \vspace*{2.cm} 
{\huge 
Four-Fermion Production
\vspace*{3mm}
\\
with Anomalous Couplings
\vspace*{3mm}
\\
at LEP2 and NLC\footnote{
Talk given at the XIIth International Workshop on High Energy Physics
and Quantum Field Theory, 4--10 September 1997, Samara, Russia
}
 \vspace*{2.cm}               %preprint 
}

{\large 
Jochen Biebel
}
\vspace*{0.5cm}
 
\begin{normalsize}
{\it
Deutsches Elektronen-Synchrotron DESY--Zeuthen,
\\
Platanenallee 6, D-15738 Zeuthen, Germany
}
\end{normalsize}
\end{center}
 
 \vspace*{1.5cm} %preprint 
 \vfill

%                       Abstract
\begin{abstract}
I give a short report on the semi-analytical approach to off-shell $W$
pair production.
In particular, I show the effects of irreducible background diagrams
of the {\tt CC11} class in the differential cross-section.
Further, I study the influence of potential anomalous couplings
at a future linear collider and investigate the sensitivity of the
forward-backward asymmetry to anomalous couplings.

\end{abstract}

\vfill

\section{Introduction}\label{introduction}

Since the start of LEP2 in 1996 the
production of $W$ pairs in $e^+e^-$ annihilation is
observed in the process $e^+e^-\to W^+W^-\to
4f$~\cite{Barate:1997mn,Abreu:1997sn,Acciarri:1997xc,Ackerstaff:1997kc}.
This offers a good possibility to measure the mass and the couplings
of the $W$ boson
\cite{Altarelli:1996ab,Beenakker:1996kt,Kunszt:1996km,Gounaris:1996rz}.
In this context, it is also a good probe of the standard model
predictions for non-Abelian gauge couplings and one way to find new
physics.

Additionally, the high energy and high luminosity of a future linear
collider will deliver a good basis for a precise determination of
anomalous couplings \cite{Accomando:1997wt}.
However, at a future collider, with a center-of-mass energy of 500~GeV
or more, the process of $W$ pair production is also
regarded as a large background for particle searches and should be
known as exactly as possible.

Therefore, it is inevitable to take certain corrections into
account. The finite width of the $W$ bosons~\cite{Muta:1986is},
irreducible background processes~\cite{Berends:1994pv,Bardin:1994sc},
and radiative corrections~\cite{Bohm:1988ck,Fleischer:1989kj,Denner:1990tx,Beenakker:1991sf,Dittmaier:1992np,Beenakker:1997ir,Denner:1997ia}
must be considered.

To make theoretical predictions for the cross-section, a
semi-analytical approach \cite{Bardin:1994sc,Bardin:1993bh,Bardin:1996uc}
is advantageous in two ways.
First of all, semi-analytical programs, like the Fortran program
{\tt GENTLE} \cite{Bardin:1996zz}, allow a fast and precise
prediction of total and differential cross-sections.
This can be used directly in the analysis of data\footnote{
{\tt GENTLE} was used e.~g.~in the search for anomalous couplings
\cite{Acciarri:1997xc}.}.
On the other hand, the
precise predictions of a semi-analytical calculation can be used to
test the reliability of the various existing Monte Carlo programs
\cite{Bardin:1997gc,Ohl:1996ig}.

In section \ref{secdiff}, I study the background effects for
processes of the {\tt CC11} class, which is defined by the final state
fermions:
\begin{itemize}
\item[--] {\tt CC09}: $(\mu\bar{\nu}_\mu, \bar{\tau}\nu_\tau)$
\item[--] {\tt CC10}: $(\mu\bar{\nu}_\mu, \bar{d}u)$, 
                  $(\mu\bar{\nu}_\mu, \bar{s}c)$,
                  $(\tau\bar{\nu}_\tau, \bar{d}u)$, 
                  $(\tau\bar{\nu}_\tau, \bar{s}c)$
\item[--] {\tt CC11}: $(d\bar{u}, \bar{s}c)$
\end{itemize}
and their charge conjugates.
A more detailed description of the classification scheme for four
fermion production can be found
in \cite{Bardin:1994sc}; see also \cite{Bardin:1997gc}.

The non-Abelian gauge structure of the standard model was subject of
theoretical interest for many years and a general expression for
the three-gauge boson couplings was developed in 1979
\cite{Gaemers:1979hg}.
Potential new physics was parameterized in an effective theory with
anomalous couplings \cite{Hagiwara:1987vm}.
The current limits for the couplings from the LEP2 measurement
are \cite{Saeki}:
\begin{eqnarray}
  \alpha_{B\phi}&=&0.45^{+0.56}_{-0.67}\nonumber\\
  \alpha_{W\phi}&=&0.02^{+0.16}_{-0.15}
\vphantom{\frac{m^2_W}{m^2_W}}
\label{limits}\\
  \alpha_{W}&=&0.15^{+0.27}_{-0.27}\nonumber
\end{eqnarray}
A definition for the parameters in (\ref{limits}) is given in section
\ref{anocoup}.
There, I study also the potential for measuring anomalous couplings
at a future linear collider.

%=====================================================================
%=====================================================================

\section{The Differential Cross-Section for {\tt CC11}:}
\label{secdiff}

The semi-analytical results for the total cross-section for $e^+e^-\to
W^+W^-\to 4f$ including the background of the {\tt CC11} class are
presented and discussed in detail in \cite{Bardin:1996uc}.
The differential cross-section for the {\tt CC03} process, i.~e.~only
the signal diagrams, is also given there.

Here, I will shortly sketch the results for the differential
cross-section.

I write the threefold differential cross-section at a center-of-mass
energy squared $s$ as follows:
\begin{equation}
  \frac{\mbox{d}^3\sigma}{\mbox{d}s_{1}\mbox{d}s_2\mbox{d}\cos\theta}
  \; = \; \frac{\sqrt{\lambda}}{\pi s^2} 
  \sum_{k} {\cal C}_k \cdot
               {\cal G}_k (s, s_1,s_2,\cos\theta)
\label{sigstc}
\end{equation}
with
\begin{equation}
   \lambda=\lambda(s,s_1,s_2)=s^2+s_1^2+s_2^2-2ss_1-2ss_2-2s_1s_2
\end{equation}
The invariant masses of the produced fermion pairs are $s_1$ and $s_2$.

In (\ref{sigstc}) I introduced the coefficient functions ${\cal
  C}_k$ and the kinematical functions ${\cal G}_k$.
While the ${\cal C}_k$'s are trivial and contain only the
coupling constants and the $s$-channel propagators, the
${\cal G}_k$'s describe the kinematics of the process and depend on
$\cos\theta$.
All ${\cal C}_k$'s and ${\cal G}_k$'s for the {\tt CC11} class will be
presented in~\cite{Biebel:1997aa}.

To include the effects of initial state radiation using the
structure function approach the Born cross-section is convoluted with
two structure functions $D(x)$ \cite{Kuraev:1985hb,Beenakker:1996kt}.
I have to change eq.~(\ref{sigstc}) to 
\begin{equation}
  \frac{\mbox{d}^3\sigma}{\mbox{d}s_{1}\mbox{d}s_2\mbox{d}
    \cos\theta_{\sf{lab}}}
  \; = \; 
  \int\!\!\mbox{d} x_+  \int\!\!\mbox{d} x_- \, D(x_+) D(x_-) \sum_i \left|
  \frac{\partial \cos\!\theta_i}{\partial
  \cos\!\theta_{\sf{lab}}}\right| 
  \frac{\mbox{d}^3\sigma}{\mbox{d}s_1\mbox{d}s_2,
    \mbox{d}\cos\theta_i}.
\label{angdiffqed}
\end{equation}

Because of the radiation of photons, the electron and the positron
change their energy and momentum.
This leads to a Lorentz boost and the scattering
angle in the center-of-mass system of the $W$ bosons is shifted
compared to the scattering angle measured in the detector.
In the calculation I correct this by a transformation of the
angles.
Since the transformation is not always unique, the number of solutions
depends on $x_-$ and $x_+$.
This is reflected in the sum over $i$.

\begin{figure}[thb]
\centerline{\psfig{figure=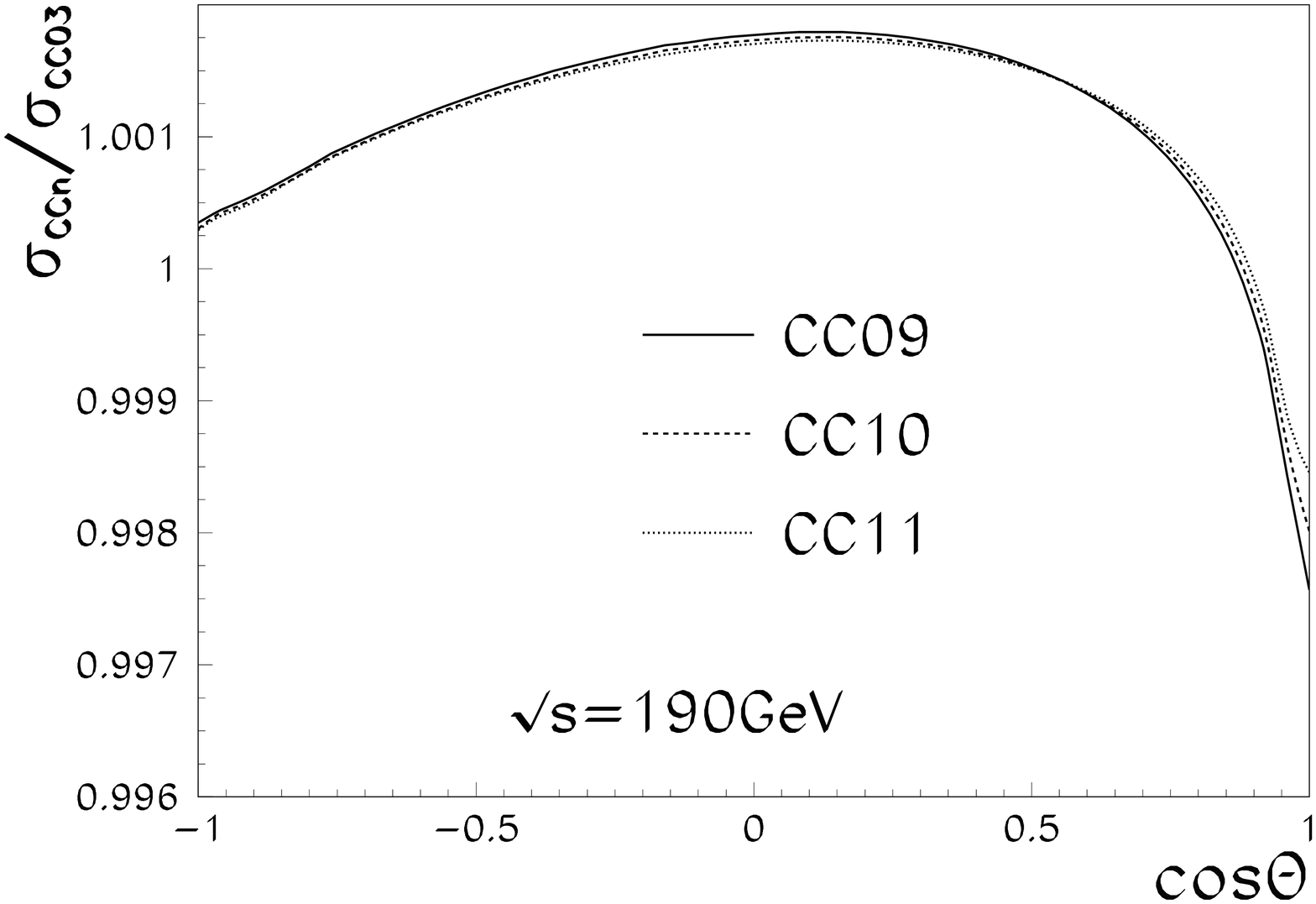,width=16cm}}
\caption{\label{diffcross}The ratio of background to signal processes
  in Born approximation.}  
\end{figure}

Note, that for a bin-wise integration of (\ref{angdiffqed}) the
relatively complicated Jacobean can be
substituted by a transformation of the bin limits.
This allows an analytical integration over $\cos\theta$.

As an application I study the different background
effects for the various final states of the {\tt CC11} class in
fig.~\ref{diffcross}.
They are shown for a center-of-mass energy of 190~GeV as the ratio of
the cross-section calculated for signal plus background diagrams over
the cross-section of signal diagrams only.

Although the final state fermions have different coupling constants
and there are even different numbers of diagrams for the {\tt CC09},
{\tt CC10}, and {\tt CC11} processes, the effects due to the
irreducible background are almost the same.

%=====================================================================
%=====================================================================

\section{Anomalous Couplings}\label{anocoup}

Since 1979 there was much attention to the subject of anomalous
couplings, see for example
\cite{Gaemers:1979hg,Hagiwara:1987vm,Bilchak:1984ur,Hagiwara:1993ck,Diehl:1994br,Jegerlehner:1994zp,Berends:1995dn,Gounaris:1996uw}.
As a recent overview \cite{Gounaris:1996rz} might be useful.

To introduce anomalous couplings into the calculation I add three
terms to the standard model Lagrangian.
The new terms  are operators of  dimension 6 and conserve both
{\sf C} {\em and} {\sf P}.
They are \cite{Gounaris:1996rz}:
%%%%%%%%%%%%%%%%%%%%%%%%%%%%%%%%%%%%%%%%%%%%%%%%%%%%%%%%%%%%%%%%%%%%%%
\begin{equation}
  \Delta{\cal L}  = \mbox{} g'\frac{\alpha_{B\phi}}{m^2_W}
    \left(D_\mu\Phi\right)^\dagger B^{\mu\nu}\left(D_\nu\Phi\right)
    +g\frac{\alpha_{W\phi}}{m^2_W}\left(D_\nu\Phi\right)^\dagger
    {\overrightarrow{\tau}} \cdot {\overrightarrow{W}}^{\mu\nu} 
    \left( D_{\nu}\Phi \right)
    + g\frac{\alpha_{W}}{6m^2_W}\overrightarrow{W}^\mu_\nu\cdot
\left(\overrightarrow{W}^\nu_\rho
 \times \overrightarrow{W}^\rho_\mu\right) 
 \label{deltala}
\end{equation}
%%%%%%%%%%%%%%%%%%%%%%%%%%%%%%%%%%%%%%%%%%%%%%%%%%%%%%%%%%%%%%%%%%%%%%
In the unitary gauge these operators lead to the following effective
Lagrangian for the $WWV$ vertex:
%%%%%%%%%%%%%%%%%%%%%%%%%%%%%%%%%%%%%%%%%%%%%%%%%%%%%%%%%%%%%%%%%%%%%%
\begin{equation}
  {\cal  L}^{WWV}_{\mbox{eff}}  =  {\sf i} g_{WWV} \left[\:
    \vphantom{\frac{\lambda_V}{m^2_W}}
    g^V_1 \left(W^+_{\mu\nu}W^{-\mu}-W^{+\mu}
    W^-_{\mu\nu}\right)V^\nu 
   +\kappa_VW^+_\mu W^-_\nu V^{\mu\nu} +
      \frac{\lambda_V}{m^2_W} W^{+\nu}_\mu W^{-\rho}_\nu V^\mu_\rho
      \right]~\nonumber,
  \label{anolag}
\end{equation}
%%%%%%%%%%%%%%%%%%%%%%%%%%%%%%%%%%%%%%%%%%%%%%%%%%%%%%%%%%%%%%%%%%%%%%
where $V$ can be a $\gamma$ or a $Z$.

Electromagnetic gauge invariance requires $g_1^\gamma=1$.

For the $WWZ$ vertex, I add also the term
%%%%%%%%%%%%%%%%%%%%%%%%%%%%%%%%%%%%%%%%%%%%%%%%%%%%%%%%%%%%%%%%%%%%%%
\begin{equation}
  {\cal L}_Z =-\frac{e z_Z}{m^2_W} \,
  \partial_\alpha\hat{Z}_{\rho\sigma}
  \left( W^{+\alpha}\stackrel{\leftrightarrow}{\partial^\rho}W^{-\sigma}-
  W^{+\sigma}\stackrel{\leftrightarrow}{\partial^\rho}W^{-\alpha}\right)~
\end{equation}
%%%%%%%%%%%%%%%%%%%%%%%%%%%%%%%%%%%%%%%%%%%%%%%%%%%%%%%%%%%%%%%%%%%%%%
with
%%%%%%%%%%%%%%%%%%%%%%%%%%%%%%%%%%%%%%%%%%%%%%%%%%%%%%%%%%%%%%%%%%%%%%
\begin{equation}
  \hat{Z}_{\rho\sigma}=\frac{1}{2}\epsilon_{\rho\sigma\alpha\beta}
   Z^{\alpha\beta}.
\end{equation}
%%%%%%%%%%%%%%%%%%%%%%%%%%%%%%%%%%%%%%%%%%%%%%%%%%%%%%%%%%%%%%%%%%%%%%
This coupling is odd under both {\sf C} {\em and} {\sf P} transformation,
therefore it is invariant under a {\sf CP} transformation.

In the standard model, we have
$\kappa_\gamma=\kappa_Z=g_1^Z=1$ and $\lambda_\gamma=\lambda_Z=z_Z=0$.

With the couplings $\kappa_\gamma$ and $\lambda_\gamma$ the magnetic
dipole moment and the electric quadrupole moment of the $W$ are
\cite{Aronson:1969aa}:
%%%%%%%%%%%%%%%%%%%%%%%%%%%%%%%%%%%%%%%%%%%%%%%%%%%%%%%%%%%%%%%%%%%%%%
\begin{eqnarray}
  \mu_W&=&\frac{e}{2m_W}(1+\kappa_\gamma+\lambda_\gamma)\\
  q_W&=&-\frac{e}{m_W^2}(\kappa_\gamma-\lambda_\gamma)
\end{eqnarray}
%%%%%%%%%%%%%%%%%%%%%%%%%%%%%%%%%%%%%%%%%%%%%%%%%%%%%%%%%%%%%%%%%%%%%%

It is useful to choose a different set of anomalous couplings in which
all of the anomalous parameters vanish in the standard model.
This is done with the transformation~\cite{Gounaris:1996rz}:
\begin{equation}
\begin{array}{lclclcl}
  x_\gamma&=&\kappa_\gamma-1&\hspace{2cm}&
  x_Z&=&(\kappa_Z-1)\cot\theta_W-\delta_Z
\\
\\
  y_\gamma&=&\lambda_\gamma&&
  y_Z&=&\lambda_Z\cot\theta_W
\\
\\
  z_Z&=&z_Z&&
  \delta_Z&=&(g_1^Z-1)\cot\theta_W
\end{array}
\label{ourpara}
\end{equation}

\begin{figure}[thb]
\centerline{\psfig{figure=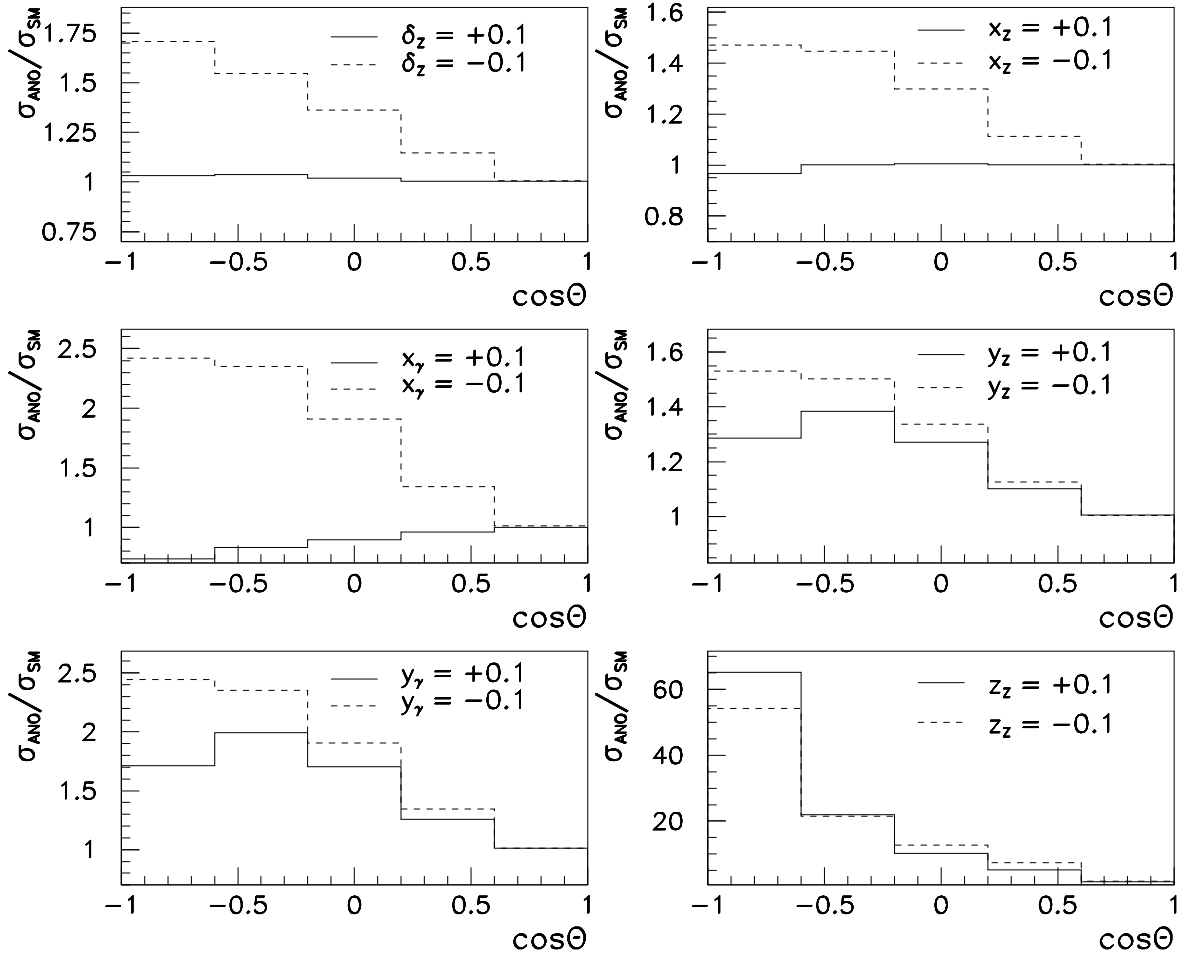,height=14cm,width=16cm}}
\caption{\label{ano5}Ratio of cross-sections at
  $\sqrt{s}=500$~GeV. Only one parameter differs from~0 in each figure.}   
\end{figure}

The effects of the anomalous couplings defined in (\ref{ourpara}) are
shown in fig.~\ref{ano5}.
To give an impression of the relative change in the cross-section the
ratio $\sigma_{\mbox{\scriptsize ANO}}/\sigma_{\mbox{\scriptsize SM}}$
is plotted.
The curves demonstrate nicely the sensitivity of
the differential cross-section to anomalous couplings for large
scattering angles, i.~e.~$\cos\theta<0$.

In addition, the plots in fig.~\ref{ano5} are suitable for comparisons
with output of Monte Carlo event generators.
For 190~GeV the {\tt GENTLE} results were in excellent agreement with
the data published in \cite{Berends:1995dn} and for an energy of
500~GeV a comparison with the Monte Carlo program {\tt WOPPER}
\cite{Anlauf:1996wq} was also successful.

\begin{figure}[thb]
\centerline{\psfig{figure=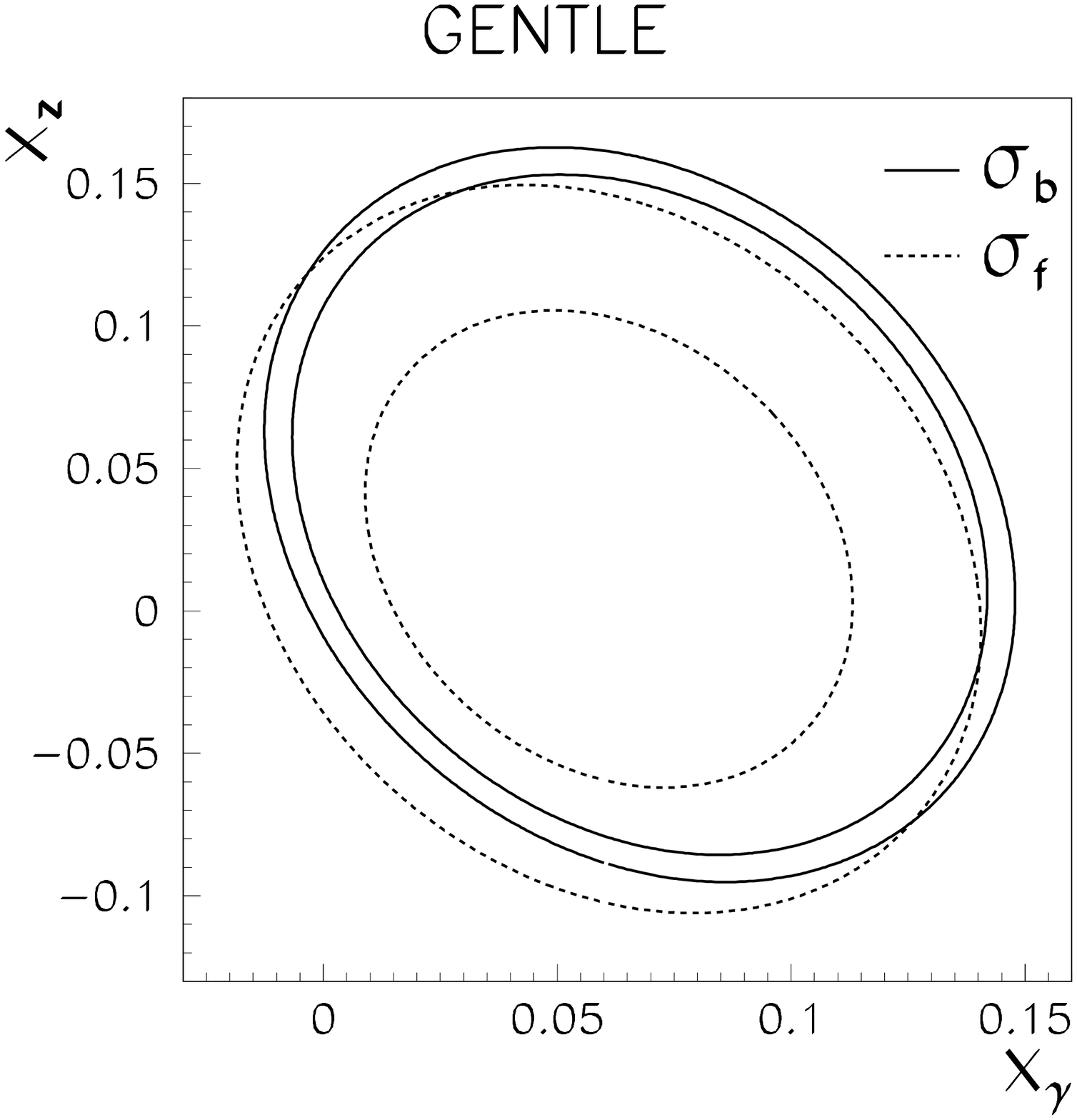,width=8cm}\psfig{figure=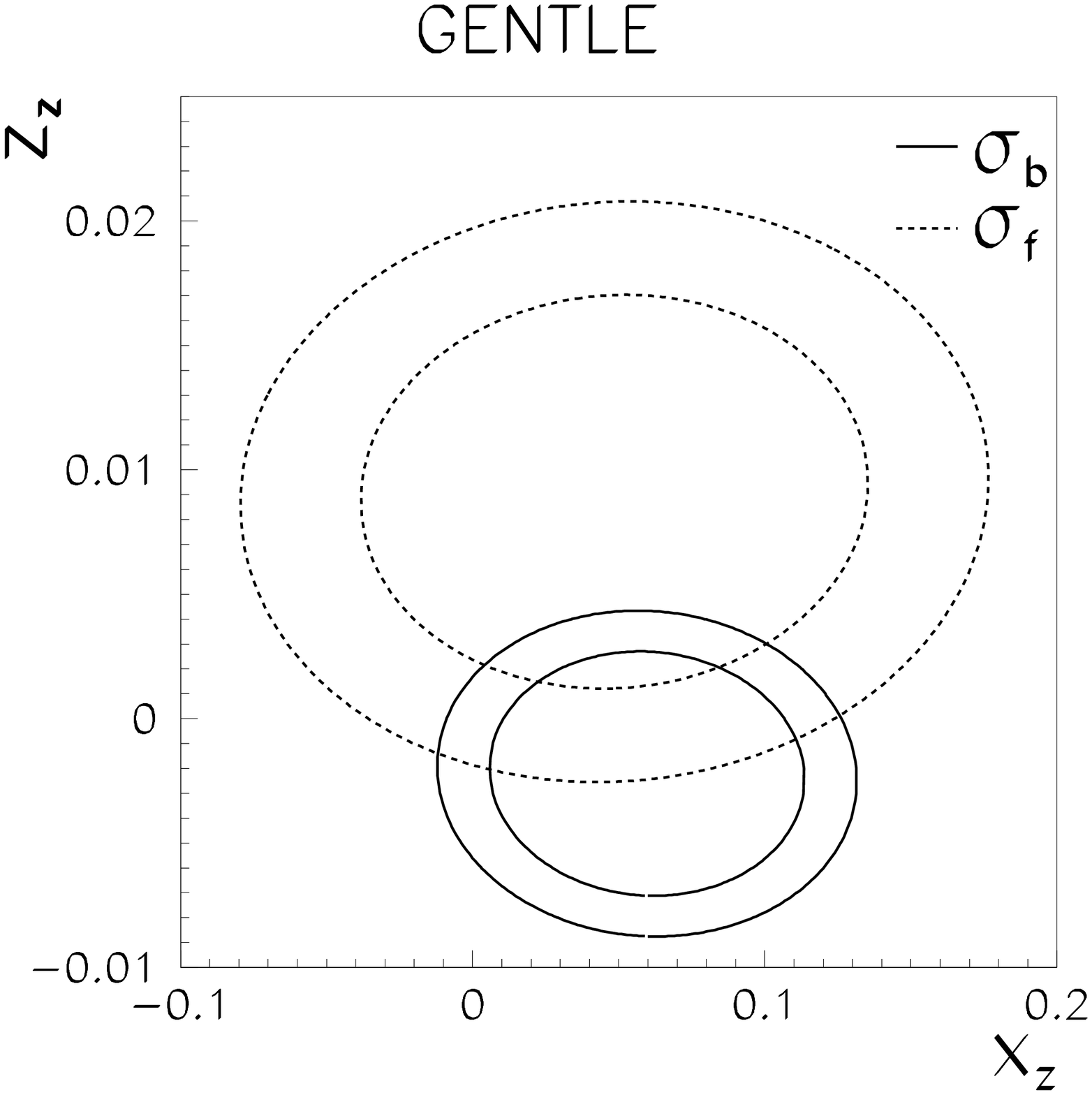,width=8cm}}
\caption{\label{rings} $1\sigma$-bounds at 500~GeV for ${\cal L}=50
  fb^{-1}$.}   
\end{figure}

Of course, it is also worthwhile and more realistic\footnote{There is
  absolutely no reason why only one of the anomalous parameters in
  (\ref{ourpara}) should appear in Nature.
  To decrease the number of new parameters it is more natural to
  require additional constraints for the operators in
  eq.~(\ref{deltala}), like in the "HISZ scenario"
  \cite{Hagiwara:1993ck,Hagiwara:1992eh}, where
  $\alpha_{B\phi}=\alpha_{W\phi}$ is assumed.} to study 
several couplings simultaneously.
In fig.~\ref{rings} I simulate the discriminative power of the
forward-backward asymmetry to detect anomalous couplings.
For a more detailed description of this analysis see
\cite{Biebel:1997ti}.
The two rings shown in each picture correspond to the cross-section
measured in the range $-1<\cos\theta<0$ (backward) and
$0<\cos\theta<1$ (forward).
The rings represent the allowed region for a pair of anomalous couplings,
if we assume that the standard model cross-section is measured.

Although the statistics for forward scattering is much higher\footnote{
At 500~GeV about 94\% of the produced $W$ bosons go into the forward
direction.}, the 
narrow rings, which give more stringent limits, correspond to the
measurement of the backward cross-section.
This proves again the strong influence of anomalous couplings in the
region of backward scattering \cite{Bilchak:1984ur}.

Fig.~\ref{rings} shows also that the allowed regions do not
overlap so much, if the parameter $z_Z$ is non-vanishing.   
It is not surprising that parity violating couplings are stronger
constrained by the forward-backward asymmetry than parity conserving
couplings.

%=====================================================================
%=====================================================================

\section{Conclusions}

I discussed some problems in connection with the use of $W$ pair
production at a linear collider in the search for new physics.
Especially in the region of backward-scattering, anomalous couplings
have a strong effect.
However, only if precise predictions
for the cross-sections exist, there is a chance to find
deviations of the standard model.

To achieve such a high precision semi-analytical programs are useful.
With the capability to produce bin-wise integrated cross-sections and
to consider the effects of anomalous couplings, {\tt GENTLE}
provides a good base for comparisons with Monte Carlo event
generators.

%=====================================================================
%=====================================================================
\bigskip

\smallskip

\noindent
{\Large\bf Acknowledgement}

\smallskip

\noindent
I would like to thank the organizers of the conference for their kind
hospitality and the warm atmosphere they provided.
Further, it is a pleasure for me to thank the other authors of
{\tt GENTLE}: D.~Bardin, D.~Lehner, A.~Leike, A.~Olchevski, and
T.~Riemann.
Especially, I am grateful to Tord Riemann for collaboration on this
topic.
Further I thank T.~Ohl for useful discussions and hints.

\end{document}